\documentclass[twocolumn,twoside,slac_two,sort&compress,numbers]{revtex4}
\usepackage{graphicx}
\usepackage{fancyhdr}
\usepackage{natbib}
\usepackage{txfonts}
\begin{document}

\title{Cosmochemistry, Cosmology, and Fundamental Constants\\
High-Resolution Spectroscopy of Damped Lyman-Alpha Systems}

%

\author{R. Quast, D. Reimers}
\affiliation{Hamburger Sternwarte, Universit\"{a}t Hamburg, D-21029 Hamburg, Germany}
\author{A. Smette, O Garcet}
\affiliation{Institut d'Astrophysique et de G\'{e}ophysique, Universit\'{e} de Li\`{e}ge, B-4000 Li\`{e}ge, Belgium}
\author{C. Ledoux}
\affiliation{European Southern Observatory, Santiago, Chile}
\author{S. Lopez}
\affiliation{Departamento de Astronom\'{i}a, Universidad de Chile, Santiago, Chile}
\author{L. Wisotzki}
\affiliation{Astrophysikalisches Institut Potsdam, D-14482 Potsdam, Germany}

\begin{abstract}
Spectroscopy of QSO absorption lines provides essential observational input for
the study of nucleosynthesis and chemical evolution of galaxies at high redshift.
But new observations may indicate that present chemical abundance data are biased
due  to deficient spectral resolution and unknown selection effects: Recent
high-resolution spectra reveal the hitherto unperceived chemical nonuniformity of
a molecule-bearing damped Lyman-alpha (DLA) system, and the still ongoing H/ESO
DLA survey produces convincing evidence for the effect of dust attenuation. We present
a revised analysis of the H$_2$-bearing DLA complex toward the QSO HE~0515--4414
showing nonuniform differential depletion of chemical elements onto dust grains,
and introduce the H/ESO DLA survey and its implications. Conclusively, we aim
at starting an unbiased chemical abundance database established on high-resolution
spectroscopic observations. New data to probe the temperature-redshift relation
predicted by standard cosmology and to test the constancy of fundamental constants
will be potential spin-offs.
\end{abstract}

\maketitle

\thispagestyle{fancy}

\section{CHEMICAL UNIFORMITY AT HIGH Z?}
In stark contrast to many interstellar lines-of-sight in the Milky Way
or the Magellanic Clouds, high-redshift DLA systems usually appear to
be chemically uniform \cite{ProchaskaJ_2003}. But despite their chemical
uniformity, for any redshift the observed metallicities are different by
up to two orders of magnitude \cite{ProchaskaGWCD_2003}.

The known DLA systems make a heterogeneous population involving very
different physical environments -- a very unfavorable precondition for
tracing nucleosythesis and the chemical evolution of galaxies. But the
physical environments of molecular clouds are less diverse. In fact, for
the known H$_2$-bearing DLA systems the metallicity-redshift distribution
may exhibit less variation than for the regular DLA systems
\cite{CurranWMC_2004}.

Molecular hydrogen is detected in DLA components exhibiting high particle
densities and low kinetic temperatures \cite{PetitjeanSL_2002,LedouxPS_2003}.
The metallicity of DLA systems is usually calculated by averaging ad hoc
radial velocity intervals, i.e. by averaging several absorption components.
But the depletion of chemical elements in H$_2$-bearing components may
surpass the average by up to one order of magnitude \cite{PetitjeanSL_2002,
LedouxPS_2003}. Moreover, narrow components arising from cold gas are blurred
in the usually complex absorption profiles.

In fact, high-resolution (55\,000) and exceptionally high signal-to-noise
(90-140) spectra of the H$_2$-bearing DLA system toward
the QSO HE~0515--4414 indicate the hitherto unperceived chemical nonuniformity
of individual metal absorption profile components, similar to the interstellar
lines-of-sight intersecting Galactic warm disk and halo clouds (Figs.~1, 2). In
addition, for the H$_2$-bearing components the calculated \cite{VladiloG_2004}
fraction of iron in dust of 98 percent is close to that of Galactic cold disk
gas.
Is the DLA system toward HE~0515--4414 just a rare case, or are compact
high-metallicity dust clouds systematically missed in present spectroscopic
observations due to insufficient resolution or low signal-to-noise ratio?

\begin{figure*}
\includegraphics{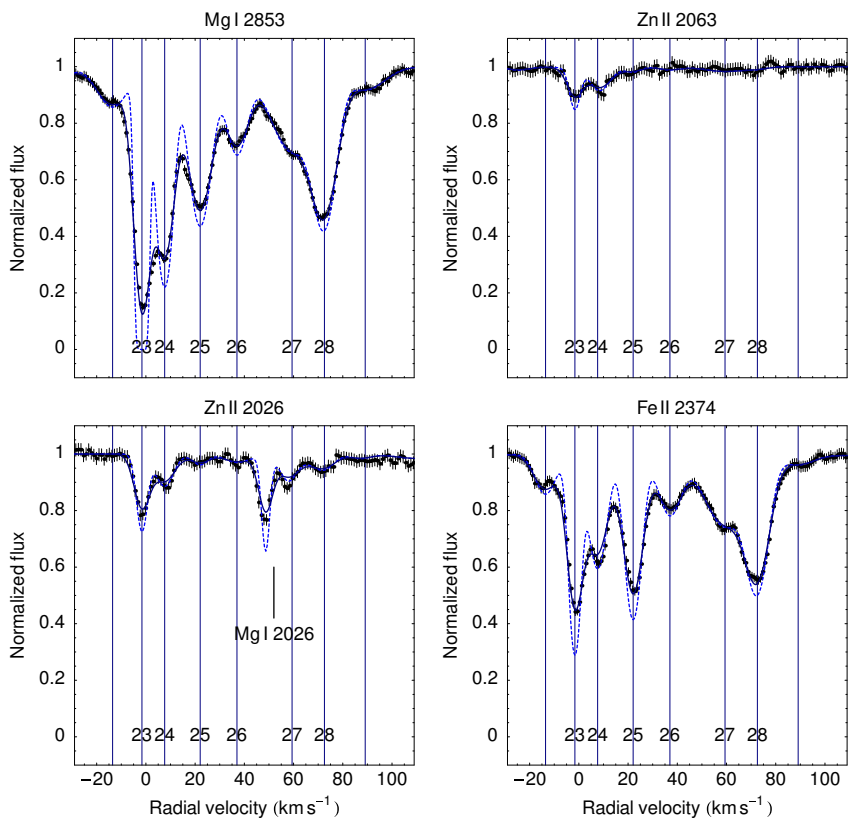}
\caption{Selected metal absorption profiles associated with the DLA system
toward HE~0515--4414. The blue curves indicate the optimized profile
decomposition and its deconvolution. Individual components are labeled by
numbers 23-28. Note the different optical depths of Fe\,\textsc{ii} and
Zn\,\textsc{ii} for the H$_2$-bearing componets 23/24 and component 28.
Component 23 also exhibits rare neutral species (S\,\textsc{i},
Si\,\textsc{i}, Fe\,\textsc{i}). Radial velocity zero corresponds to
the redshift $z=1.1508$.} \label{f1}
\end{figure*}

\begin{figure}[t]
\includegraphics{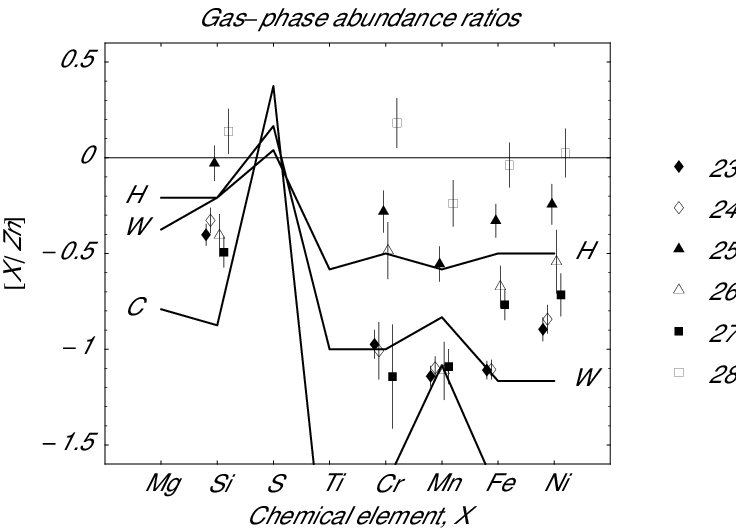}
\caption{Evidence for chemical nonuniformity at high redshift. Gas-phase
abundance ratios (relative to solar ratios) for the absorption components
shown in Fig.~1 compared with typical values found for Galactic cold (C)
and warm (W) disk and halo (H) clouds \cite{WeltyLBHY_2001}. Since the
volatile element Zn is only mildly depleted, the abundance ratios reflect
the differential depletion of chemical elements into dust. Note that Cr,
Mn, Fe, and Ni are strongly depleted in the H$_2$-bearing components
23/24 but appear essentially undepleted in component 28.} \label{f2}
\end{figure}

\section{FAINT QSOS OBSCURED BY DUST?}

The possibility that present surveys of DLA systems are affected by dust is
an ongoing concern: If the extinction of DLA absorbers is high enough,
optical surveys will miss the QSOs behind them. However, present samples
of DLA absorbers toward radio-selected and SDSS QSOs do not indicate any
distinct selection bias due to dust \cite{EllisonCRP_2004,
MurphyM_LiskeJ_2004}.

In contrats, the new H/ESO survey of DLA systems toward a complete
subsample of 182 HE QSOs produces convincing evidence for the effect of
dust attenuation: Four new DLA systems were discovered toward the bright
half of the QSO subsample, but 14 were detected toward the faint. The
probability for both numbers being drawn from the same Poissonian is less
than 2.5 percent (Fig.~3).
Do the DLA systems toward the fainter QSOs exhibit more dust than those
toward the bright?

\begin{figure}[t]
\includegraphics{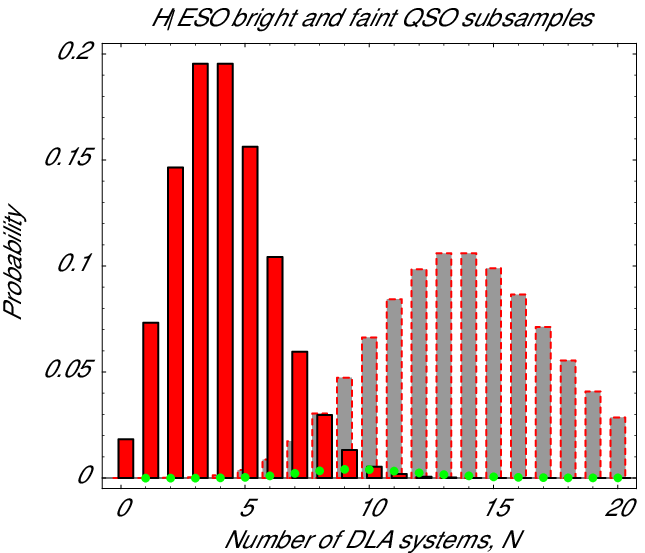}
\caption{Evidence for attenuation of fainter QSOs by dust. The number
of DLA systems detected toward the bright (faint) half of 182 HE QSOs
is 4 (14). The absorption paths of both subsamples exhibit almost the same
length. Two Poissonians with means 4 and 14 are marked by, respectively,
the red and gray vertical bars. The probability for both numbers being drawn
from the same Poissonian with mean $N$ is indicated by green dots.} \label{f3}
\end{figure}

\section{FUTURE AIMS}

In order to study the possible biases and selection effects involved with
duts, we aim at starting a sound abundance database established on
high-resolution and high signal-to-noise spectroscopy of the complete
sample of DLA systems discovered by the H/ESO survey. The database will
be extremely useful for studying the physical conditions in DLA systems,
and may shed some light on the star formation history of the universe and
the problem of missing metals \cite{WolfePG_2003,WolfeGP_2003}.

As spin-off product, newly detected C\,\textsc{i} and C\,\textsc{ii}
fine-structure absorption lines may be used to test the
temperature-redshift relation predicted by the standard big-bang
cosmology. A further potential spin-off will be a homogenous sample
of Fe\,\textsc{ii} lines suitable for probing the hypothetical variation
of the fine-structure constant $\alpha$ by means of the regression 
many multiplet method \cite{LevshakovS_2004,QuastRL_2004}.

\begin{acknowledgments}
This research is based on observations made with the ESO Very Large
Telescope under Programme ID~066.A-0212 and hs been funded by the
Verbundforschung of the BMBF/DLR under Grant No. 50\,OR9911\,1.
\end{acknowledgments}

\bibliography{abbrev,astron,rq}

\end{document}